\documentclass[amsmath,twocolumn,prb]{revtex4-1}
\RequirePackage{varioref}
\RequirePackage{amsmath}
\RequirePackage{amssymb}
\RequirePackage{bm}
\RequirePackage{graphicx}
\newcommand{\refeq}[1]{Eq.~\eqref{#1}}
\newcommand{\reffig}[1]{Fig.~\ref{#1}}
\newcommand{\etal}{et~al.}
\renewcommand{\vec}[1]{\ensuremath{\text{\textbf{#1}}}} 

\renewcommand{\Re}[1]{\ensuremath{
    \text{Re}\left\{#1\right\}}}

\newcommand{\mathcommand}[3][0]{\newcommand{#2}[#1]{\ensuremath{#3}}}
\newcommand{\ham}[1][]{\ensuremath{{H}_{\text{#1}}}} 

\newcommand{\nodag}{{\phantom{\dag}}} 

\newcommand{\tsc}[2]{{#1}_{\textsc{#2}}} 
\newcommand{\be}{\begin{equation}}
\newcommand{\ee}{\end{equation}}
\mathcommand{\erf}{{\rm erf}}
\mathcommand{\erfc}{{\rm erfc}}
\mathcommand{\sech}{{\rm sech}}
\mathcommand{\csch}{{\rm csch}}

\mathcommand[1]{\floor}{\left\lfloor #1 \right\rfloor}
\mathcommand[1]{\ceil}{\left\lceil #1 \right\rceil}
\mathcommand[1]{\smallket}{|#1\rangle}
\mathcommand[1]{\smallbra}{\langle #1|}
\mathcommand[1]{\bigket}{\bigl|#1\bigr\rangle}
\mathcommand[1]{\bigbra}{\bigl\langle #1\bigr|}
\mathcommand[1]{\biggket}{\biggl|#1\biggr\rangle}
\mathcommand[1]{\biggbra}{\biggl\langle #1\biggr|}
\mathcommand[1]{\ket}{\left| #1 \right\rangle}  
\mathcommand[1]{\lket}{\bigl| #1 \bigr)}  
\mathcommand[1]{\bra}{\left\langle #1 \right|}  

\mathcommand[1]{\cvec}{\overline{#1}}
\mathcommand{\trmat}{\cvec{\text{Tr}}}
\begin{document}
\title{Hyperfine-Induced Decay in Triple Quantum Dots}
\author{Thaddeus D. Ladd}
\affiliation{HRL Laboratories, LLC, 3011 Malibu Canyon Rd., Malibu, CA 90265-4797, USA}
\date\today
\begin{abstract}
We analyze the effects of hyperfine interactions on coherent control experiments in triple quantum dots.  By exploiting Hamiltonian symmetries and the SU(3) structure of the triple-dot system under pseudo-exchange and longitudinal hyperfine couplings, we provide analytic formulae for the hyperfine decay of triple-dot Rabi and dephasing experiments.
\end{abstract}
\maketitle
\section{Introduction}

Recent experiments in triple quantum dots\cite{laird_triple,nrc_triple,medford} show increased coherent control abilities over previous experiments in double dots\cite{petta,laird_double,reilly_nuclei,yacoby_universal,maune_nature}.  In both cases, one factor limiting coherence is the hyperfine interaction to substrate nuclear spins.  In this brief report, we derive simple expressions for the form of hyperfine-induced decay in simple triple-dot experiments, and indicate how these experiments may be used to elucidate the total volume of the quantum dot wavefunctions.  The formalism we introduce may also impact the development of composite control sequences for mitigating hyperfine-induced pulse errors.

In our analysis, we treat the nuclei as providing classical random fields.  In an applied magnetic field sufficiently large to energetically forbid electron-nuclear flip-flops, the hyperfine Hamiltonian may be well approximated as $\ham[HF]=\sum_j B_j S_j^z$, where $S_j^z$ is the $z$-component of the spin operator for spin $j$, with $S=1/2$.  Here $z$ refers to the direction of the applied magnetic field. The hyperfine fields $B_j$ are summations over all nuclear spins within each quantum dot volume, with appropriate
coupling constants\cite{coishloss,maune_nature}.  These sums are treated as classical random variables drawn from a zero-mean Gaussian distribution of variance $\langle B_j^2\rangle = \sigma_j^2$. The average value of $\sigma_j$ across three dots, which we notate $\tsc\sigma{hf}$, provides our scaling throughout this analysis.  In real time units, $\tsc\sigma{hf}^{-1}\sim T_2^*\sim 10$~ns for GaAs-based dots\cite{petta,laird_double} and about 300~ns for Si-based dots\cite{maune_nature}.  Nuclear fluctuations are driven by nuclear dipole-dipole kinetics which are very slow in comparison to typical control and measurement timescales\cite{reilly_nuclei}.

Coherent control of the spins is achieved by pseudo-exchange between dots 1 and 2, or between dots 2 and 3.  Exchange operators are written in terms of spin operators as $E_{jk} = \vec{S}_j\cdot\vec{S}_k$.  The exchange terms of the Hamiltonian may be modulated via voltage pulses.

Naively, the analysis of hyperfine-averaged dynamics under various exchange pulses may be performed by constructing the full Hamiltonian in an 8-dimensional basis and then averaging over the 3 independent random variables $B_j$.  Employing three symmetries, we reduce this problem to a one-dimensional integral with a good analytic approximation; the resulting expressions are then particularly useful for curve-fitting, allowing the estimation of quantum dot volumes via hyperfine parameters.

\section{SU(3) Description of Triple-Dot}

\begin{figure}
\includegraphics[width=\columnwidth]{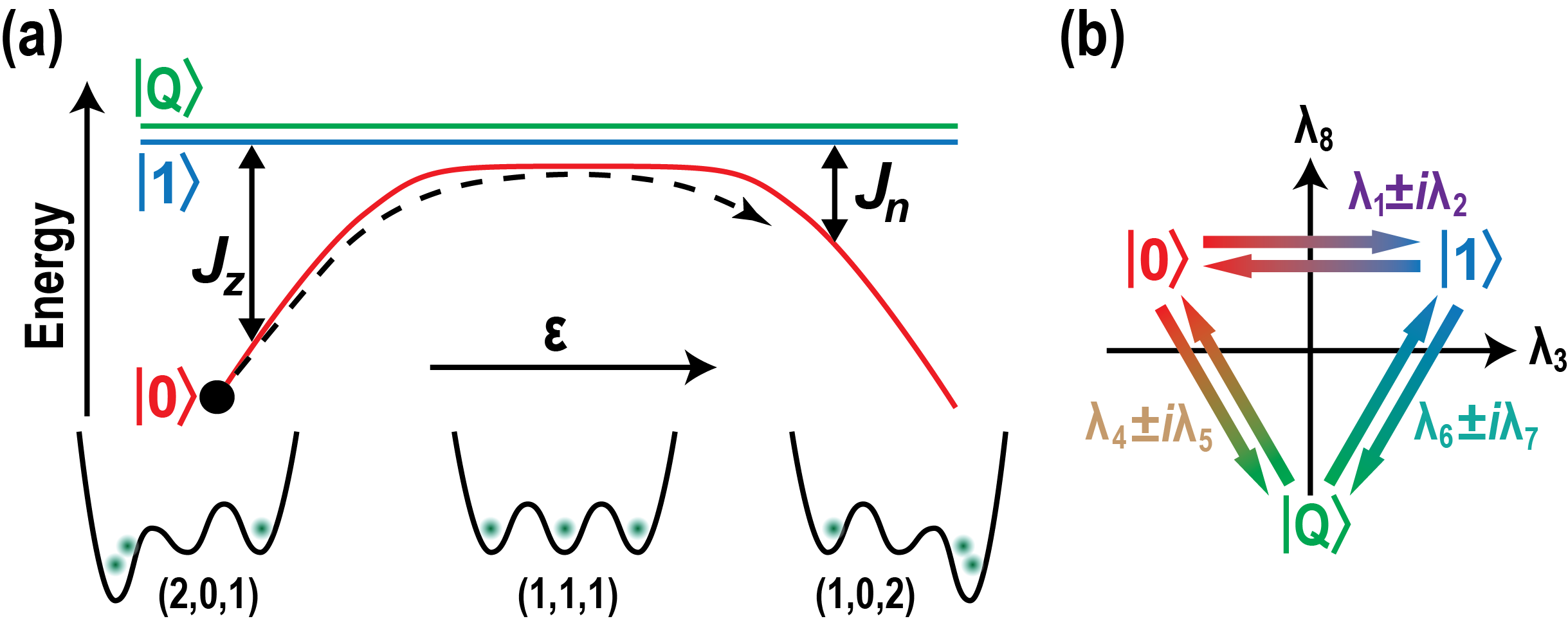}
\caption{(a) Schematic of energy levels of triple-dot system, above a sketch of triple-dot energy potentials.  The black dot and black dashed arrow indicate the initialization and pulsing of a triple-dot Rabi or dephasing experiment.
(b) SU(3) structure of triple dot system.
}
\label{triple_dot_SU3}
\end{figure}

The first important symmetry of both \ham[HF]\ and the exchange operators is that they all conserve the $z$-projection of the total angular momentum of the three spins.  It follows that the Hamiltonian may be block-diagonalized into four blocks corresponding to quantum number $m_z=\pm 3/2, \pm 1/2$.  Since current experiments initialize into a spin-singlet state, the two $m_z = \pm 3/2$ states are never occupied, and hence we will discuss them no further.  The remaining six states are divided into two manifolds, which we notate $\Uparrow$ and $\Downarrow$ for $m_z=1/2$ and $m_z=-1/2$, respectively.  We refer to this degree of freedom as a gauge qubit.

A convenient set of $E_{12}$ eigenstates in the $\Uparrow$ manifold are
\be
\begin{split}
\ket{0\Uparrow} &=
\bigket{\bar{D}_{+1/2}'}
=\frac{\ket{\uparrow\downarrow\uparrow}-\ket{\downarrow\uparrow\uparrow}}{\sqrt{2}}
\\
\ket{1\Uparrow}
&= \bigket{\bar{D}_{+1/2}}
=\frac{\ket{\uparrow\downarrow\uparrow}+\ket{\downarrow\uparrow\uparrow}}{\sqrt{6}}
-\sqrt{\frac{2}{3}}\ket{\uparrow\uparrow\downarrow}
\\
\bigket{Q\Uparrow} &=\frac{\ket{\uparrow\downarrow\uparrow}+\ket{\downarrow\uparrow\uparrow}+\ket{\uparrow\uparrow\downarrow}}{\sqrt{3}}.
\end{split}
\label{upbasis}
\ee
The states are notated according to logical states of a qubit in a decoherence free subsystem (DFS)\cite{divincenzo_dfs,kempe_dfs}.  The labels $\bigket{\bar{D}_{\pm 1/2}},$ etc. are those of Refs.~\onlinecite{laird_triple,nrc_triple}, provided here for comparison.  The $\Downarrow$ manifold may be found by flipping all three spins in the $\Uparrow$ manifold.  This choice of states is physically motivated by the ability to initialize into the ground state of the exchange operator $E_{12}$ by preparing a ground state in the (2,0,1) charge configuration, as shown in \reffig{triple_dot_SU3}(a).  This ground state is chosen as logical 0.  The state labeled $Q$ has the critical feature of being fully symmetric and therefore equally unaffected by exchange on any pair of dots; this is therefore not useful for coherent control and is referred to as a ``leaked" state.  The other logical qubit state is the only remaining orthogonal state within the given gauge.

Figure \ref{triple_dot_SU3}(a) sketches energy levels in only one of the gauge manifolds, for simplicity; the other one possesses the same structure.  We therefore drop the $\Uparrow,\Downarrow$ distinction from the state labels.  More complete diagrams may be found in Refs.~\onlinecite{laird_triple,nrc_triple}.
On the left of this figure, corresponding to negative detuning parameter $\epsilon$, the three dots have the $(2,0,1)$ charge configuration, and as $\epsilon$ is increased, hybridization with the $(1,1,1)$ charge configuration results in pseudo-exchange of magnitude $J_z$, indicated as the energy difference between the red and blue lines, corresponding respectively to the $\ket{0}$ state $\ket{1}$ states.  The states $\ket{0},\ket{1},$ and $\ket{Q}$ are energy eigenstates only far to the left of this energy diagram.  The subscript $z$ is chosen for $J_z$ since the Hamiltonian term $J_zE_{12}$ splits the qubit states, and is therefore logically analogous to the Pauli operator $\sigma_z$ on the logical qubit. Far to the right of the diagram, pseudo-exchange between dots 2 and 3 has magnitude $J_n$.

The choice of $n$ subscript on $J_n$ is made evident by considering the group structure for a single gauge. All operations of exchange control and hyperfine drift within one gauge are described by the group SU(3).  Using the states in \refeq{upbasis} in the order given, a natural selection of generators are the Gell-Mann matrices in each basis, traditionally labeled $\lambda_j$ for $j=1\ldots 8$.  These matrices may be interpreted as raising and lowering through a triangle of basis states as diagrammed in \reffig{triple_dot_SU3}(b).  The exchange operators are independent of gauge, and may be written in terms of the SU(3) generators as
\begin{align}
E_{12}=&-\frac{\lambda_8}{2\sqrt{3}}-\frac{\lambda_3}{2}
    \label{J12}\\
E_{23}=&-\frac{\lambda_8}{2\sqrt{3}}-\frac{\lambda_3\cos\phi+\lambda_1\sin\phi}{2},
    \label{J23}
\end{align}
where $\phi=120^\circ$.
This representation makes evident the effect of each term for the DFS qubit Bloch sphere, defined by the projections to $\lambda_1,\lambda_2,\lambda_3$.  The $\lambda_8$ in each term commutes with these generators; it provides a relative phase to the $\ket{Q}$ state.  The $E_{12}$ term provides rotations about the $\lambda_3$ or $\hat{z}$ axis, and hence we ascribe it the magnitude $J_z$.   The $E_{23}$ term provides rotations about the $\hat{n}=[\sin\phi,0,\cos\phi]$ axis, and hence we ascribe it the magnitude $J_n$.   The exchange Hamiltonian $\ham[ex]=J_zE_{12}+J_nE_{23}$ is diagonalized by the unitary $U_2(\eta)=\exp(-i\eta\lambda_2/2)$, where $\eta = \tan^{-1}[J_n\sin\phi/(J_n\cos\phi+J_z)]$, providing the eigenstates $\bigket{\Delta_{\pm 1/2}}$ and $\bigket{\Delta_{\pm 1/2}'}$ of Ref.~\onlinecite{laird_triple}.

\newcommand{\DQ}[1]{\Delta_{\overline{#1}}}
\newcommand{\SQ}[1]{\sigma_{\overline{#1}}}

In our SU(3) picture, the hyperfine Hamiltonian is written
\be
\label{Hhf}
\tsc{H}{hf} = \pm U_7^\nodag(\beta)\biggl(\frac{\Delta_{12}}{2}\lambda_1+\frac{\DQ{12}}{\sqrt{3}}\lambda_8\biggr)
U_7^\dag(\beta),
\ee
where $U_7(\beta)=\exp(-i\lambda_7\beta/2)$ for $\beta=\pi-\tan^{-1}\sqrt{8}\approx 109.5^\circ$.  The hyperfine differences are $\Delta_{12} = B_1-B_2,$ and $\DQ{12} = B_3-(B_1+B_2)/{2}$.
We will also use the linear combinations $\Delta_{23}$ and $\DQ{23}$, defined via permutation of the subscripted dot labels.  These hyperfine field differences have variances and covariance
\begin{align}
\langle \Delta_{12}^2\rangle &= \sigma_{12}^2 = \sigma_1^2+\sigma_2^2,\\
\langle \DQ{12}^2\rangle &= \SQ{12}^2 = \sigma_3^2 + \frac{\sigma_1^2+\sigma_1^2}{4},\\
\langle \Delta_{12}\DQ{12} \rangle &= C_{12} = \frac{\sigma_1^2-\sigma_2^2}{2}.
\end{align}
The action of these hyperfine terms can be assessed using \reffig{triple_dot_SU3}(b).  First, $U_7(\beta)$ mixes the $\ket{1\Uparrow}$ and $\ket{Q\Uparrow}$ states
into $\ket{T_{12}^0\uparrow}$ and $\ket{T_{12}^{+1}\downarrow}$, where $\ket{T_{12}^m}$ is the $J^z=m$ projection of the $J=1$ triplet for spins 1 and 2. These states are degenerate with respect to $E_{12}$, and hence $[\lambda_7,E_{12}]=0$, which provides our second key symmetry allowing analytic diagonalization.  Following this fixed SU(3) rotation, the hyperfine Hamiltonian is highly reminiscent of the double-dot singlet-triplet case: $\Delta_{12}$ causes only $\hat{x}$-axis rotations of the double-dot singlet-triplet qubit, while $\DQ{12}$ provides a relative phase to the polarized triplet state.

The final symmetry we employ is that the exchange interaction is independent of the two gauge manifolds,  $\Uparrow$ and $\Downarrow$, while \refeq{Hhf} exactly flips sign between them. We briefly note that this points to hyperfine decoupling schemes in which the total electron spin is rapidly flipped, perhaps with electron-spin-resonance-based \mbox{$\pi$-pulses}; such schemes would have no direct effect on exchange interactions.  For the purposes of our present analysis, if an experiment averages over $\Delta_{12}$ and $\DQ{12}$, and if the probability distributions for these variables are symmetric about zero, then the average dynamics will be the same in each manifold.  This allows us to drop the $\Uparrow,\Downarrow$ distinction and work with a single SU(3) system.

There is one further energy term in the system which we have not addressed, that of a global magnetic field, either experimentally applied or due to average hyperfine terms.  A global field splits the energy of the gauge manifolds, but plays no further role; it is of course the invariance to this field and its fluctuations to which the nomenclature ``decoherence-free" applies\cite{divincenzo_dfs,kempe_dfs}.

\section{Rabi and Dephasing Experiments}

The algebra we have introduced enables an analytic description of two of the simplest experiments which may be performed in a pulsed triple-dot system. In both experiments, the singlet $\ket{0}$ is prepared in dots 1 and 2 [see \reffig{triple_dot_SU3}(a)] with the gauge state typically random.  (High magnetic-field experiments may allow for initialization into a single gauge state).  For a Rabi-like experiment, a bias sweep then quickly reduces the magnitude of $J_z$ while increasing the magnitude of $J_n$.  Free evolution with $J_n>0$ then occurs for time $t$, after which $J_z$ is quickly ramped up as $J_n$ is quickly ramped down. The probability of returning to the initial singlet state is observed via Pauli blockade on the (1,1,1)/(2,0,1) charge transition and subsequent charge-state detection\cite{laird_triple,nrc_triple,medford}.  A dephasing experiment in a triple dot is the same experiment, except $J_n$ is chosen to be a very small value, optimally zero.

In Ref.~\onlinecite{laird_triple}, the triple-dot Hamiltonian is diagonalized for nonzero $J_n$ and $J_z$, but with vanishing hyperfine.  Here, we diagonalize for a single nonvanishing exchange term and nonvanishing hyperfine.  For the experiments described, the Hamiltonian to be diagonalized is $H=\tsc{H}{hf}+J_nE_{23}$.  We begin by diagonalizing $E_{23}$ via the rotation $U_2(\phi)=\exp(-i\phi\lambda_2/2)$, since $U_2^\dag(\phi) E_{23} U_2^\nodag(\phi)=E_{12}$.  In general, this rotation may be considered to permute the dot labels; the rotated hyperfine Hamiltonian may therefore be taken as \refeq{Hhf} with the $\{1,2\}$ subscripts permuted to $\{2,3\}$.  Stated another way, the problem of initializing as a singlet in dots $1$ and $2$ and pulsing $J_n$ is equivalent to the problem of initializing as a singlet in dots $2$ and $3$ and pulsing $J_z$, except that we must swap the labels of dots 1 and 3 in the hyperfine parameters.  As we have seen, we may then apply $U_7(\beta)$ to block-diagonalize $H$, and we are left with diagonalization of a single-qubit Hamiltonian.  A final $\lambda_2$ rotation by angle $\theta = -\tan^{-1}\Delta_{23}/{J_n}$ completes the diagonalization:
\begin{multline}
U_2^\dag(\theta)U_7^\dag(\beta)U_2^\dag(\phi) [\tsc{H}{hf}+J_nE_{23}] U_2^\nodag(\phi) U_7^\nodag(\beta) U_2^\nodag(\theta)
\\
=
-\frac{\sqrt{J_n^2+\Delta_{23}^2}}{2}\lambda_{3}
-\frac{J_n-2\DQ{23}}{2\sqrt{3}}\lambda_{8}.
\end{multline}
\begin{figure}
\includegraphics[width=\columnwidth]{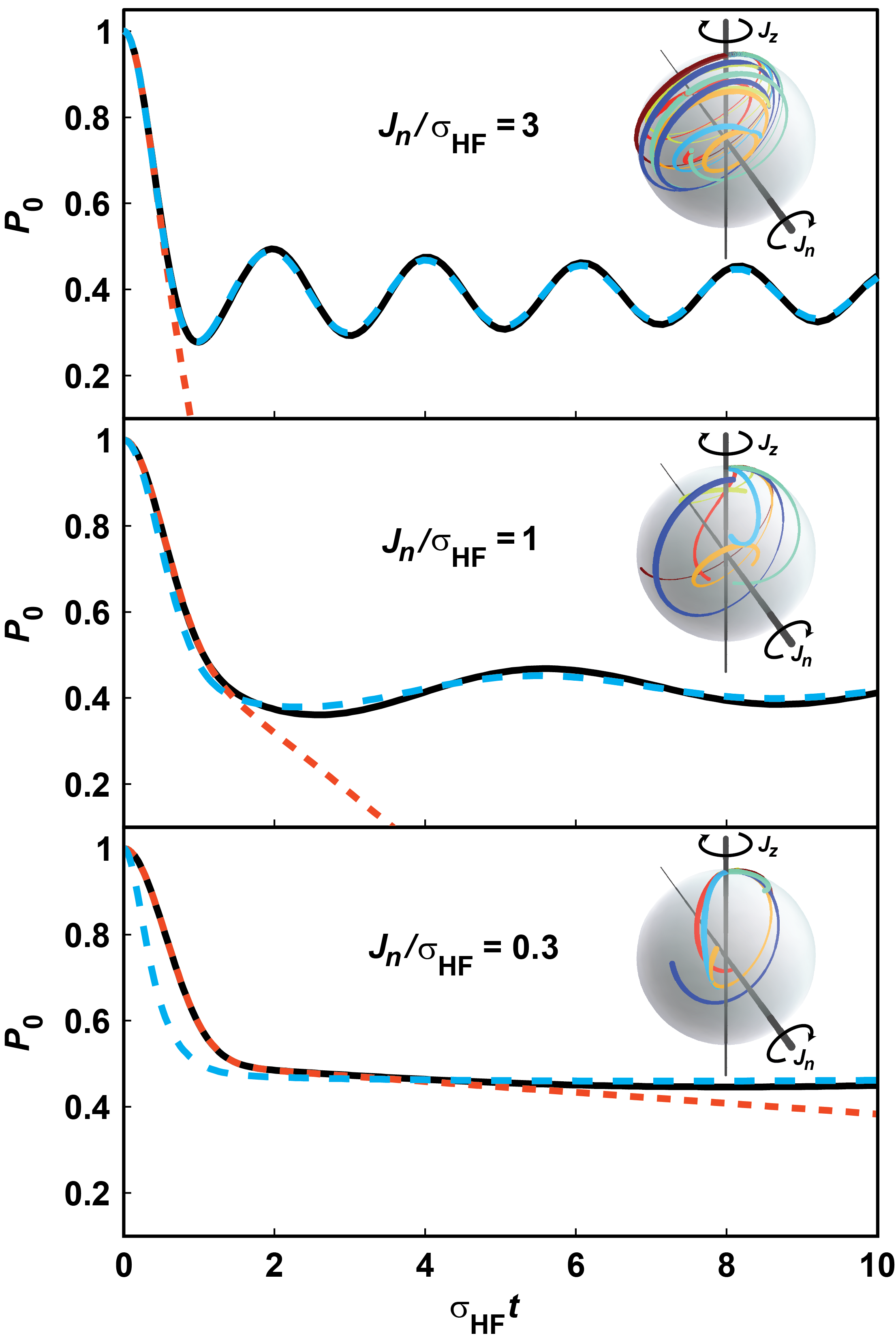}
\caption{
Numeric example of $P_0(t)$ for three values of $J_n.$  The example values for the hyperfine deviations for the three dots used for this plot are $\sigma_j = (j/2)\tsc\sigma{hf}.$  The black curve shows numeric integration of \refeq{main}, featuring no approximation relative to full $8\times 8$ diagonalization and three-dimensional averaging. The blue dashed line shows the high-$J_n$ approximation [Eqs.~(\ref{hiJ1}-\ref{hiJphi})].   The red line (short dashes) shows the low-$J_n$ approximation [\refeq{lowJ}].  For each $J_n$ a Bloch sphere is shown, in which the three cardinal directions are the expectation value of the first three Gell-Mann matrices $\lambda_1, \lambda_2$, and  $\lambda_3$.  Hence the surface is the space of DFS qubit states, while the interior corresponds to leakage.  The 8 colored trajectories represent the unitary evolution under 8 different hyperfine fields, sampled from the 3 independent Gaussian distributions of the 3 dots.  $P_0(t)$ is the average projection of these trajectories onto the $z$ axis, from which all trajectories begin.}
\label{threeJ}
\end{figure}
We then calculate the probability $P_0(t) =
\langle|\!\bra{0}e^{-i(\tsc{H}{hf}+J_nE_{23})t}\ket{0}\!|^2\rangle$
%
by averaging $\Delta_{23}$ and $\DQ{23}$ over their joint Gaussian distribution; the integral over $\DQ{23}$ is trivial.  The final result may be expressed in terms of two one-dimensional integrals over $\Delta_{23}$ as
\begin{multline}
\label{main}
P_0(t) = \\
\frac{1}{2}
-\frac{1}{4}I_1(t)+\frac{1}{4}e^{-(\sigma_{23}^2\SQ{23}^2-C_{23}^2)t^2/2\sigma_{23}^2}
    \Re{(1+e^{iJ_nt})I_2(t)}.
\end{multline}
The first integral is
\be
I_1(t)=\frac{(yu)^2}{4\sqrt{2\pi}}\int_{-\infty}^\infty dx
    \ \text{sinc}^2\biggl(\frac{\sqrt{x^2+y^2}}{2}u\biggr)
    e^{-x^2/2},
\ee
where $u=\sigma_{23}t$, $x=\Delta_{23}/\sigma_{23}$, and $y=J_n/\sigma_{23}$.  In terms of the same variables, the second integral is
\begin{multline}
I_2(t) = \frac{1}{\sqrt{2\pi}}\int_{-\infty}^\infty dx\biggl[\cos(wxu)-i\frac{x\sin(wxu)}{\sqrt{x^2+y^2}}\biggr]\times\\
e^{-x^2/2+i(\sqrt{x^2+y^2}-y)u/2},
\end{multline}
where $w=C_{23}/\sigma_{23}^2$.
These integrals resemble those occurring in the analysis of double-dot experiments\cite{coishloss}.  Examples are plotted as the black lines in \reffig{threeJ}.

Fortunately, the Rabi and dephasing experiments described above function in one of two limits.  The Rabi experiment operates in the high $y=J_{n}/\sigma_{23}$ limit.  For $y\rightarrow\infty$, $I_{1}\rightarrow 1$ and $I_2(t)\rightarrow\exp[-(C_{23}t/\sigma_{23})^2/2]$ resulting in the simple expression
\be
\label{infJ}
\lim_{J_n\rightarrow\infty} P_0(t) = \frac{3}{8}+\frac{\cos(J_nt)}{8}+\frac{1+\cos(J_nt)}{4}e^{-(\SQ{23}t)^2/2}.
\ee
In contrast to double-dot Rabi experiments\cite{petta,maune_nature}, this expression features an offset and oscillatory term decaying with effective $T_2^*$ of $\sqrt{2}/\SQ{23}$, independent of $J_n$.  The first order corrections to the integrals are expressed in terms of the width function
\be
A(t,\xi,w) = \frac{1}{\sqrt{1+[(\sigma_{23}^2t/\xi)(1+2w)]^2}}
\ee
as
\begin{multline}
\label{hiJ1}
I_1(t) \approx F\biggl(\frac{J_n}{\sqrt{2}\sigma_{23}}\biggr)\times\\
    \frac{1-\sqrt{A(t,J_n,0)}\cos\{J_nt+(1/2)\cos^{-1}A(t,J_n,0)\}}{2},
\end{multline}
where $F(x)=\sqrt{2\pi}x\exp{x^2}\text{erfc}(x)$,
and
\begin{multline}
\label{hiJ2}
\Re{(1+e^{iJ_nt})I_2(t)} \approx
    \frac{A^{3/2}(t,2J_n,0)}{A(t,2J_n,C_{23}/\sigma_{23}^2)}\times\\
    e^{-A^2(t,2J_n,0)C_{23}^2t^2/2\sigma_{23}^2}
    \{\cos[\Phi(t)]+\cos[Jt+\Phi(t)]\},
\end{multline}
where
\begin{multline}
\label{hiJphi}
\Phi(t) = \frac{3}{2}\cos^{-1}A(t,2J_n,0) -\cos^{-1}A(t,2J_n,C_{23}/\sigma_{23}^2)\\
-(C_{23}/\sigma_z^2)^2[1-A^2(t,2J_n,0)(t)].
\end{multline}
These functions describe an additional long-tailed decay with timescale $J_n/\sigma_{23}^2$ and associated phase modulation, similar to Rabi experiments in the double-dot system\cite{coishloss,petta,maune_nature}.  These long tails are often obscured by the effects of charge noise\cite{taylor_chargenoise}.

The dephasing experiment operates in the low $J_{n}/\sigma_{23}$ limit.  In the limit $J_n=0$,
\be
\lim_{J_n\rightarrow 0} P_0(t) = \frac{1}{2}\biggl(1+e^{-(\sigma_{12}t)^2/2}\biggr),
\ee
exactly the same as in the double-dot case\cite{coishloss,petta,laird_double} since the third dot plays no role.  The situation is quite different from the double-dot case at finite $J_n$; a low-order approximation to the integrals $I_1(t)$ and $I_2(t)$ yields
\begin{widetext}
\begin{multline}
P_0(t) = \frac{1}{2}\biggl[
    1+\cos\biggl(\frac{J_nt}{2}\biggr)e^{-(\SQ{23}t)^2/2}
    \biggr]
+\frac{J_n^2t}{16\sigma_{23}}\biggl\{
    2\frac{1-\exp[-(\sigma_{23}t)^2/2]}{\sigma_{23} t}
    -\sqrt{2\pi}\erf\biggl(\frac{\sigma_{23} t}{\sqrt{2}}\biggr)
    +
    \cos\biggl(\frac{J_nt}{2}\biggr)\times\\
       e^{-(\SQ{23}^2\sigma_{23}^2-C_{23}^2)(\sigma_{23}t)^2/2}
       \biggl\{
            2e^{-(\sigma_2^2t/\sigma_{23})^2/2}\frac{\exp(C_{23}t^2)-1}{\sigma_{23}t}
            -\sqrt{2\pi}\biggl(\frac{\sigma_2}{\sigma_{23}}\biggr)^2\biggl[
                \erf\biggl(\frac{\sigma_2^2t}{\sqrt{2}\sigma_{23}}\biggr)
               +\erf\biggl(\frac{\sigma_3^2t}{\sqrt{2}\sigma_{23}}\biggr)\biggr]
               \biggr\}\biggr\}+O\biggl(\frac{J_n^2}{\sigma_{23}^2}\biggr).
\label{lowJ}
\end{multline}
\end{widetext}
This approximation as well as the high-$J$ approximation are compared to exact integrals in \reffig{threeJ}.

\section{Discussion}

Several intuitive observations may be extracted from our analytic results.  First, if one seeks to measure the volume of electronic wavefunctions in a triple-dot system via hyperfine effects, one needs enough information to explicitly obtain $\sigma_1,\sigma_2$, and $\sigma_3$, in addition to the material-dependent hyperfine coupling constants and isotopic content.  Our expressions show that the hyperfine-decay of Rabi oscillations provides $\SQ{23}$, while the dephasing experiment provides $\sigma_{12}$.  These two decay periods are sufficient to extract $\sigma_3^2$ explicitly.  If signal-to-noise and charge-noise concerns allow it, $\sigma_{23}$ may in principle be extracted from either experiment via curve-fitting, although a more reliable measure is to physically prepare the initial singlet via the (1,0,2) charge state and pulse $J_z$, providing $\SQ{12}$ and $\sigma_{23}$ via the dominant decay in Rabi and dephasing experiments.  Note that similar experiments using singlet-triplet control in double-dots do not allow extraction of hyperfine parameters of individual dots, only the variance of their difference\cite{coishloss,laird_double,maune_nature}.

Another important difference between triple-dot Rabi and double-dot Rabi is the ability to distinguish noise in the exchange parameter, $J$, typically caused by charge noise, from that of hyperfine noise. Equation~\ref{infJ} shows two oscillatory terms of comparable magnitude, one showing hyperfine decay and one not.  Fluctuations of $J$ (perhaps due to detuning fluctuations) in averaged experiments would result in decay of both these terms, but hyperfine decay is only active on one of them, allowing one to distinguish the different effects within a single dataset.  Yet further information about the noise characteristics is available from more complex pulsed experiments, such as triple-dot composite pulsing and spin-echo-like experiments.  The SU(3)-based analysis we provide here may assist the analysis of hyperfine effects in these more complex experiments, as well.

\acknowledgements

Bryan Fong, Brett Maune, and Andy Hunter provided valuable discussions.  Sponsored by United States Department of Defense.  The views and conclusions contained in this document are those of the authors and should not be interpreted as representing the official policies, either expressly or implied, of the United States Department of Defense or the U.S. Government.  Approved for public release, distribution unlimited.

\end{document}